\documentclass[%
aip,
pop,%
amsmath,amssymb,
reprint,%
]{revtex4-1}

\usepackage{graphicx}
\usepackage{epspdfconversion}
\usepackage{dcolumn}
\usepackage{bm}
\begin{document}

\preprint{AIP/123-QED}

\title{Identification of time scales of the violation of the
Stokes-Einstein relation in Yukawa liquids}

\author{Zahra Ghannad}
 \altaffiliation{z.ghannad@alzahra.ac.ir}

\affiliation{%
 Department of Physics, Alzahra University, P. O. Box 19938-93973, Tehran, Iran
}%

\date{\today}

\begin{abstract}
We investigate the origin of the violation of the Stokes-Einstein (SE) relation in two-dimensional Yukawa liquids. Using comprehensive molecular dynamics simulations, we identify the time scales supporting the violation of the SE relation $D\propto (\eta/T)^{-1}$, where $D$ is the self-diffusion coefficient and $\eta$ is the shear viscosity.
We first compute the self-intermediate scattering function $F_s(k,t)$, the non-Gaussian parameter $\alpha_2$, and the autocorrelation function of the shear stress $C_{\eta}(t)$. The timescales obtained from
these functions are included the structural relaxation time $\tau_{\alpha}$, the peak time of the non-Gaussian parameter $\tau_{\alpha_2}$, and the shear stress relaxation time $\tau_{\eta}$. We find that $\tau_{\eta}$ is coupled with $D$ for all temperatures indicating the SE preservation, however, $\tau_{\alpha}$ and $\tau_{\alpha_2}$ are decoupled with $D$ at low temperatures indicating the SE violation. Surprisingly, we find that the origins of this violation are related to the non-exponential behavior of the autocorrelation function of the shear stress and non-Gaussian behavior of the distribution function of particle displacements. These results confirm dynamic heterogeneity that occurs in two-dimensional Yukawa liquids that reflects the presence of regions in which dust particles move faster than the rest when the liquid cools to below the phase transition temperature. 
\end{abstract}

\maketitle


\section{\label{sec:level1}Introduction}
The well-known Stokes-Einstein (SE) relation relates the diffusion coefficient $D$ of a Brownian particle immersed in a liquid to the shear viscosity $\eta$ of a liquid at temperature $T$. For three-dimension liquids, it is given by~\cite{Einstein,Hansen}
\begin{equation}
D=\frac{k_BT}{c\pi\eta R}
\label{eq1}.
\end{equation}
Here, $R$ is the effective radius of the Brownian particle, $k_B$ is the Boltzman constant, and $c$ is a constant that depends on the
 boundary condition at the particle surface. For two-dimensional (2D) liquids, due to the different dimensionality of $\eta$ in 2D, the SE relation is given by~\cite{Liu2006,Pan2005}
\begin{equation}
D=\frac{k_BT}{c'\pi\eta}
\label{eq2}.
\end{equation}
Generally, the SE relation can be written as $D\eta/T= constant$, and any deviations from it reveal a SE violation.

The violation of the Stokes-Einstein relation is a significant anomaly that occurs in liquids~\cite{Dubey2019,Kawasaki2017,Kawasaki2019,Kawasaki,Becker2006,Jeong2010,Koddermann2008,Tsimpanogiannis2019,Alonso2007,Mallamace2010,Kob1997}. A two-dimensional Yukawa liquid (2DYL)  is one of these liquids for which the SE violation has been reported when the liquid is cooled to near its phase transition temperature $T_{\rm{m}}$~\cite{Liu2006}. 2DYL is a liquid dusty plasma composed of charged dust particles that are strongly coupled together and interact through the Yukawa potential,  i.e., $\phi(r)= Q^2\rm{exp}(-\it{r}/\lambda_D)/\rm{4}\pi \epsilon_0\it{r}$, where $\lambda_D$ is the Debye shielding length and $Q$ is the dust charge ~\cite{Yukawa1935,Konpka2000}. Due to the electric field in the plasma sheath,
dust particles can be confined and floated in a monolayer, with an ignorable out-of-plane motion to form a 2D dusty plasma liquid~\cite{Ghannad2019,Feng2016,Wang2018,Feng2017}.

The violation of the SE relation in liquids is attributed to the dynamic heterogeneity, that is, the presence of regions in which particles move faster than the rest, i.e., they are more mobile than other particles.
Although the dynamic heterogeneity is cited as a commonly proposed reason for the SE violation in literature~\cite{Tarjus1995,Pan2017,Sengupta2013}, the true origin of the violation remains unclear.
In this study, we quantify the dynamic heterogeneity in 2DYL to gain more insight into the origin of the SE violation. We provide time scales that are the signatures of the dynamic heterogeneity. We investigate the time of structural relaxation, the non-Gaussian parameter, and the non-exponentiality of the autocorrelation function of the shear stress and we show how each leads to violation of the SE relation. This paper with identifying time scales enhances our understanding of the SE violation mechanism and reports a significant advance in understanding the SE violation.

The paper is organized as follows. In Section \ref{Model}, we describe the fundamental features of the simulation technique to mimic 2DYL. In Section \ref{Transport}, we compute transport coefficients including the self-diffusion coefficient and the shear viscosity.  In Section \ref{Time scales}, we compute correlation functions including the self-part of the intermediate scattering function and the autocorrelation function of the shear stress to identify time scales and interpret the physical origins of the SE violation. In Section \ref{Conclusions}, we present the conclusions.

\section{\label{Model}MODEL AND SIMULATION TECHNIQUE}

 We performed extensive molecular dynamics (MD) simulations to mimic 2DYL~\cite{Frenkel}. We integrated the equation of the motion $m\ddot{\boldsymbol{\mathrm{r}}}_i=-\nabla\Sigma_j\phi_{ij}$ for $N$= 1024 dust particles, where $m$ is the mass of the dust particle, and $\phi_{ij}$ is the Yukawa pair interaction potential~\cite{Yukawa1935,Konpka2000}. An equilibrium Yukawa system can be characterized by two dimensionless parameters~\cite{Ohta2000,Hartmann2019}. The first is the screening parameter $\kappa=a/\lambda_D$. The second is the Coulomb coupling parameter $\Gamma=Q^2/4\pi\epsilon_0ak_BT$, where $\epsilon_0$ is the dielectric constant, $Q$ is the charge of a dust particle, $T$ is the kinetic temperature of dust particles, $a=(n\pi)^{-1/2}$ is the Wigner-Seitz radius for 2D systems, and $n$ is the surface number density of dust particles. We chose $\kappa=0.56$, which is a common value in dusty plasma experiments~\cite{Donko2006,Liu2005,Liu2006}. For this value of $\kappa$, the phase transition is at $\Gamma_{\rm{m}}\approx$ 142, so that for
$\Gamma$ $<$ $\Gamma_{\rm{m}}$($T>T_{\rm{m}}$), the 2D Yukawa system is in the liquid phase~\cite{Hartmann2005}.
 
 We applied normalized units in this work. The normalized temperature is $\Gamma^{-1}$. The time is normalized by $\omega_{pd}^{-1}$, where $\omega_{pd}=(Q^2/2\pi \epsilon_0 ma^3)^{1/2}$ is the nominal 2D dusty plasma frequency~\cite{Kalman2004}. The length is normalized by $a$, the wave number by $a^{-1}$, the shear viscosity by $nm\omega_{pd}a^2$, and the self-diffusion coefficient by $\omega_{pd}a^2$.
Initially, the particles were placed randomly into a rectangular box, and the periodic boundary condition was applied to eliminate boundary effects due to the finite size of the box. The size of the box is chosen so that the surface number density is consistent with the definition of the Wigner-Seitz radius. Thus, we chose the box with the sizes of $56.99a\times 49.08a$ so that $n\approx1/(\pi a^2)$.
The algorithm for integrating the equation motion is Verlet algorithm~\cite{Swope1982} with the integration time step of 0.037 $\omega^{-1}_{pd}$~\cite{Feng2016}, which is adequately small to conserve energy. A Nos$\acute{e}$-Hoover thermostat~\cite{Nose1984,Hoover1985} was used to maintain constant temperature $T$. The equilibration period was $10^5$ time steps and the length of the production runs was $10^6$ time steps.

According to a standard test, if the simulations model a canonical ensemble in thermal equilibrium, the ratio of the variance of the temperature obtained from them to the variance of the temperature in the canonical ensemble, $2\langle T\rangle^2/(dN)$($d$ is the dimension), must be equal to unity~\cite{Holian1995,Ghannad2019}. We reached a satisfactory result for this ratio, i.e., 0.997.

\section{\label{Transport}Transport properties}

To investigate the Stokes-Einstein relationship, we first need to calculate transport properties including the self-diffusion coefficient $D$ and the shear viscosity $\eta$ in a 2DYL.

\subsection{\label{Difusion}Self-diffusion coefficient}
The self-diffusion coefficient can be computed from the linear fit of the mean-squared displacement (MSD) versus time as~\cite{Maginn2019}
\begin{equation}
D=\mathrm{\lim}_{t\to\infty}\frac{\langle\vert\boldsymbol{\mathrm{r}}_i(t)-\boldsymbol{\mathrm{r}}_i(0)\vert^2\rangle}{4t},
\label{eq3}
\end{equation}
where
$\boldsymbol{\mathrm{r}}_i(t)$ is the position of particle $i$ at time $t$ and $\langle...\rangle$ denotes an ensemble average. It is important to choose the time interval to fit in a way that the motion of dust particles is diffusive and $D$ is meaningful. Generally, the MSD obeys a power-law, i.e., MSD($t$) $\propto t^{\mu}$, where for $\mu$ = 1, the motion is diffusive and for $\mu\neq1$, it is anomalous. A very short time, $\omega_{pd}t$ $\lesssim$ 5, must be excluded from the time interval of the fit because it shows a ballistic motion due to the trapping of dust particles in a cage created by neighboring dust particles (Fig.~1(a)). Also, very long times must be excluded due to the statistical error.
 We have selected 100 $<$ $\omega_{pd}t$ $<$ 1000 to sample $D$ and we can be sure that the motion of dust particles is diffusive in this regime (Fig.~1(a)).

The results of the computed values of $D$ for $\Gamma>88$ are shown in Fig.~1(b). For $\Gamma\lesssim88$, the exponent of MSD($t$) is $\mu\neq$ 1 at all times (Fig.~1(a)), which indicates anomalous diffusion, as a result, $D$ is meaningless and the SE relation cannot be examined. Therefore, in the following, we examine the SE relation for the temperature range $88<\Gamma<\Gamma_{\rm{m}}.$

\begin{figure}[!htp]

\includegraphics[width=9cm,height=6.5cm]{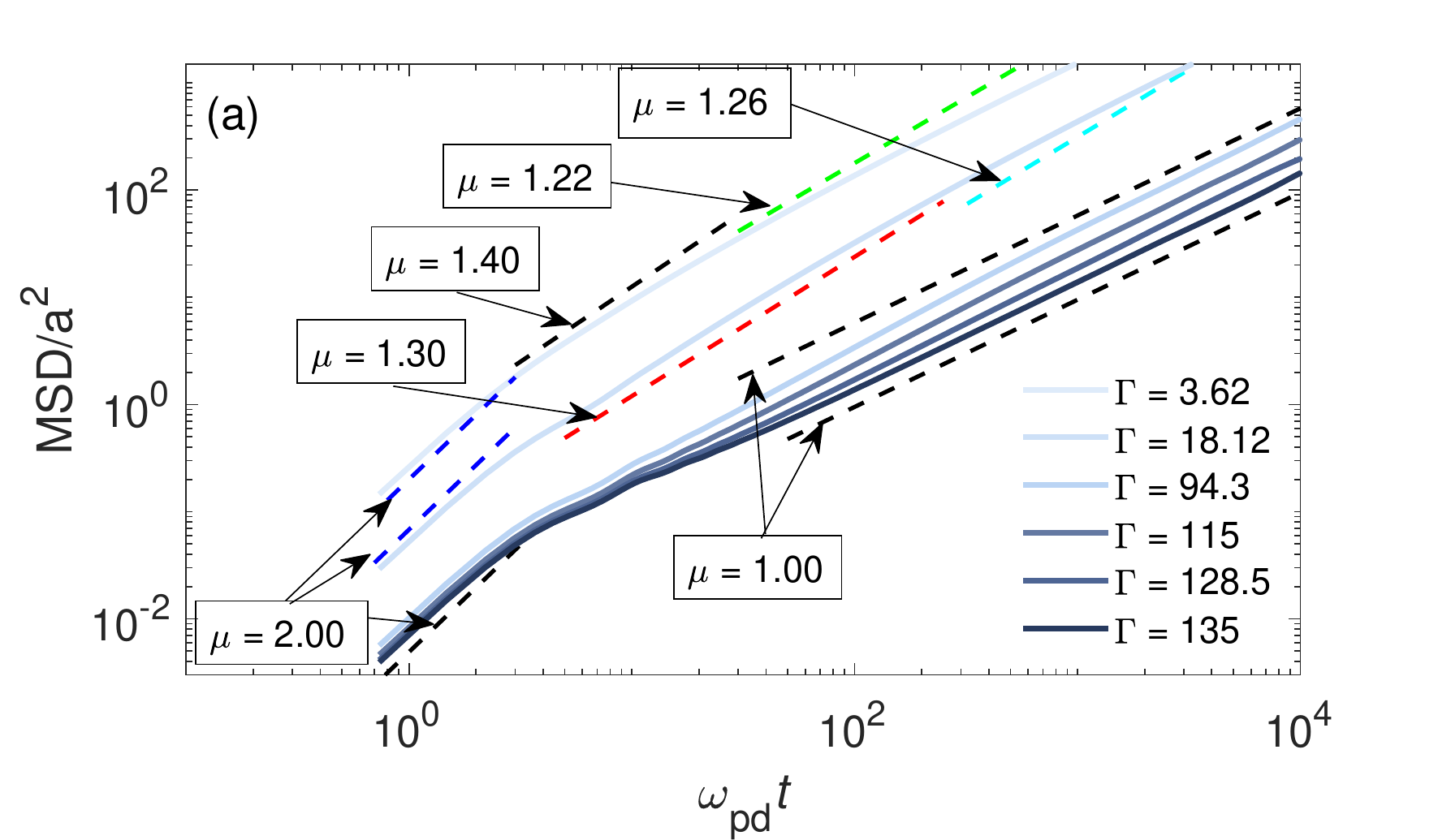}
\includegraphics[width=9cm,height=6.5cm]{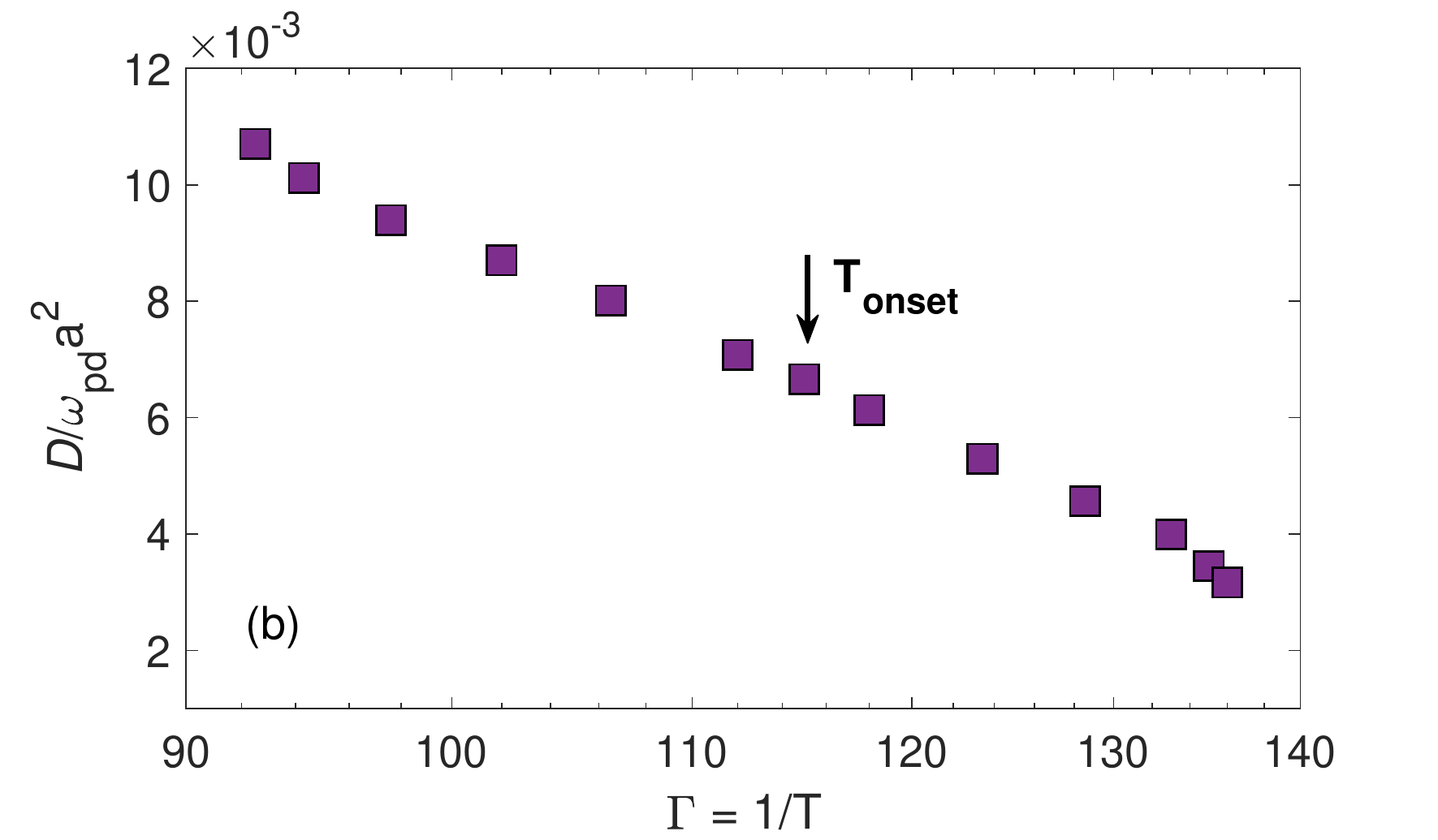}

\caption{\label{figure1} (a) Mean-squared displacements $\langle(\Delta \mathit{r}(t))^2\rangle $ as a function of time for different temperatures. For short times, the motion is ballistic, MSD($t$)$\propto t^2$. At later time,  For $\Gamma\lesssim88$, $\mu$ is not 1.00 at all times indicating anomalous diffusion, but for $\Gamma>88$, the motion is diffusive, MSD($t$)$\propto t$. 
(b) Temperature dependence of self-diffusion coefficient $D$. The error bars for $D$ are much smaller than the symbol size. $T_{\rm{onset}}$ specifies the temperature threshold in which the SE relation is violated.}
\end{figure}

\begin{figure}[!htp]

\includegraphics[width=9cm,height=6.5cm]{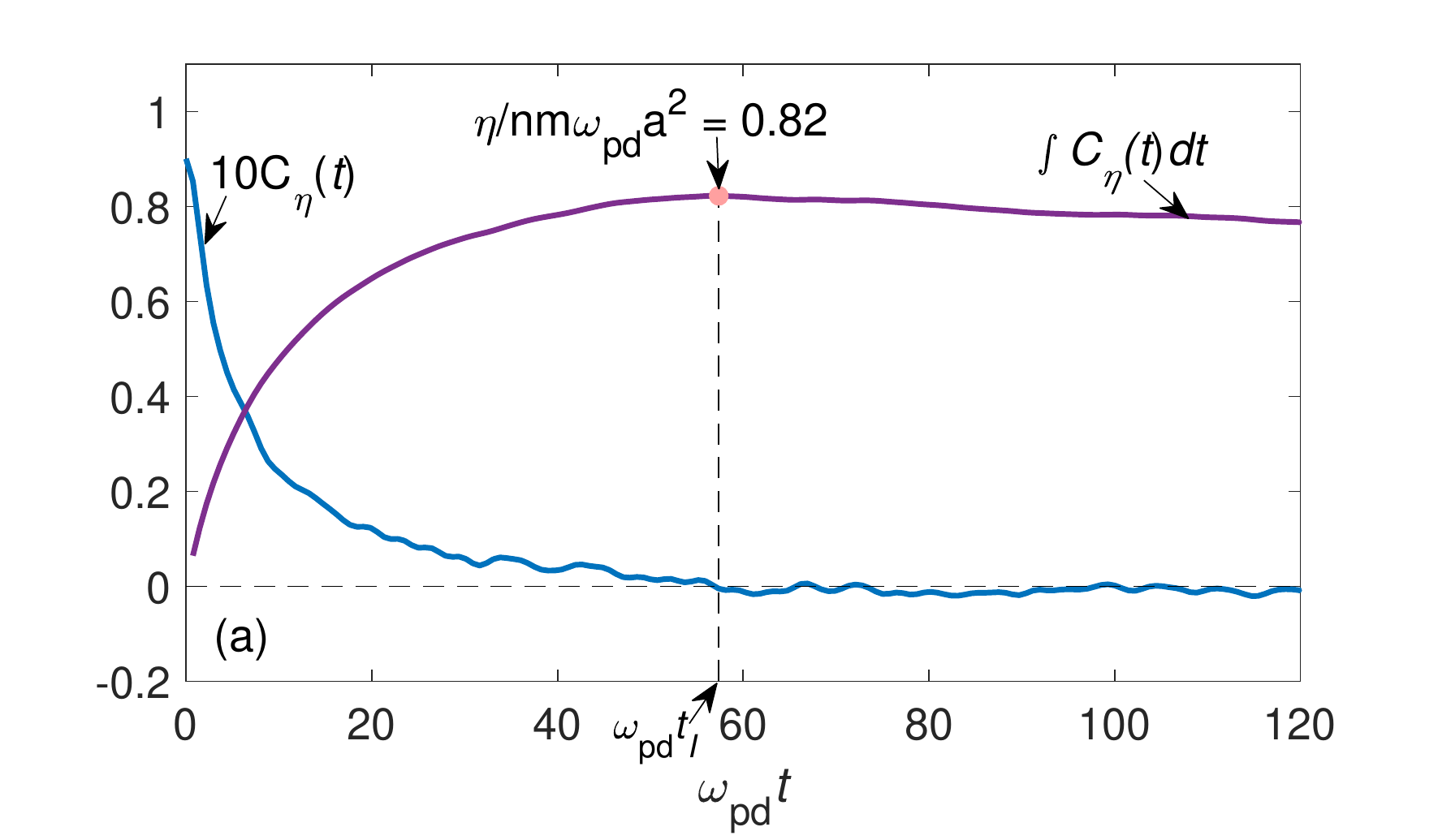}
\includegraphics[width=9cm,height=6.5cm]{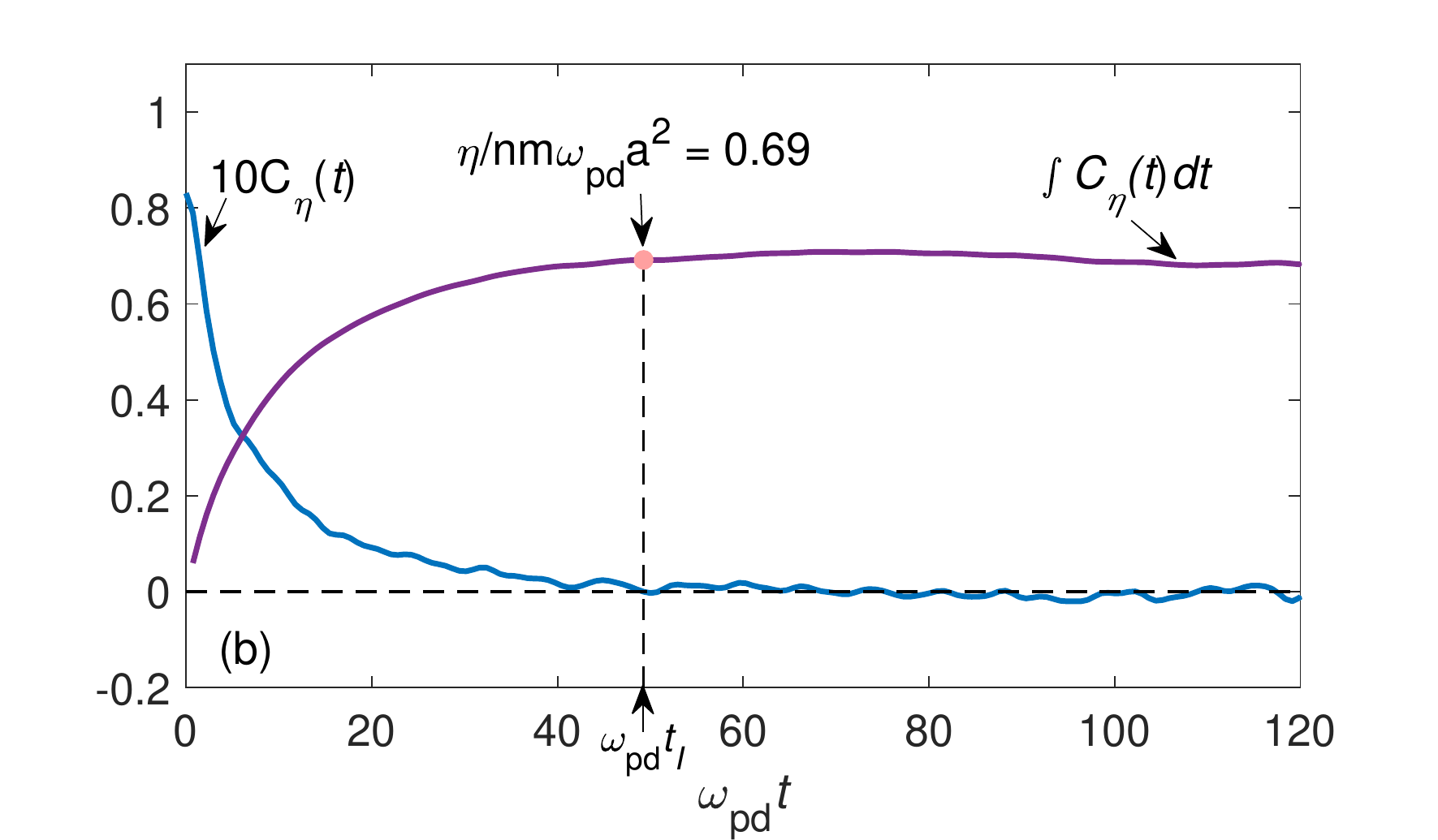}
\includegraphics[width=9cm,height=6.5cm]{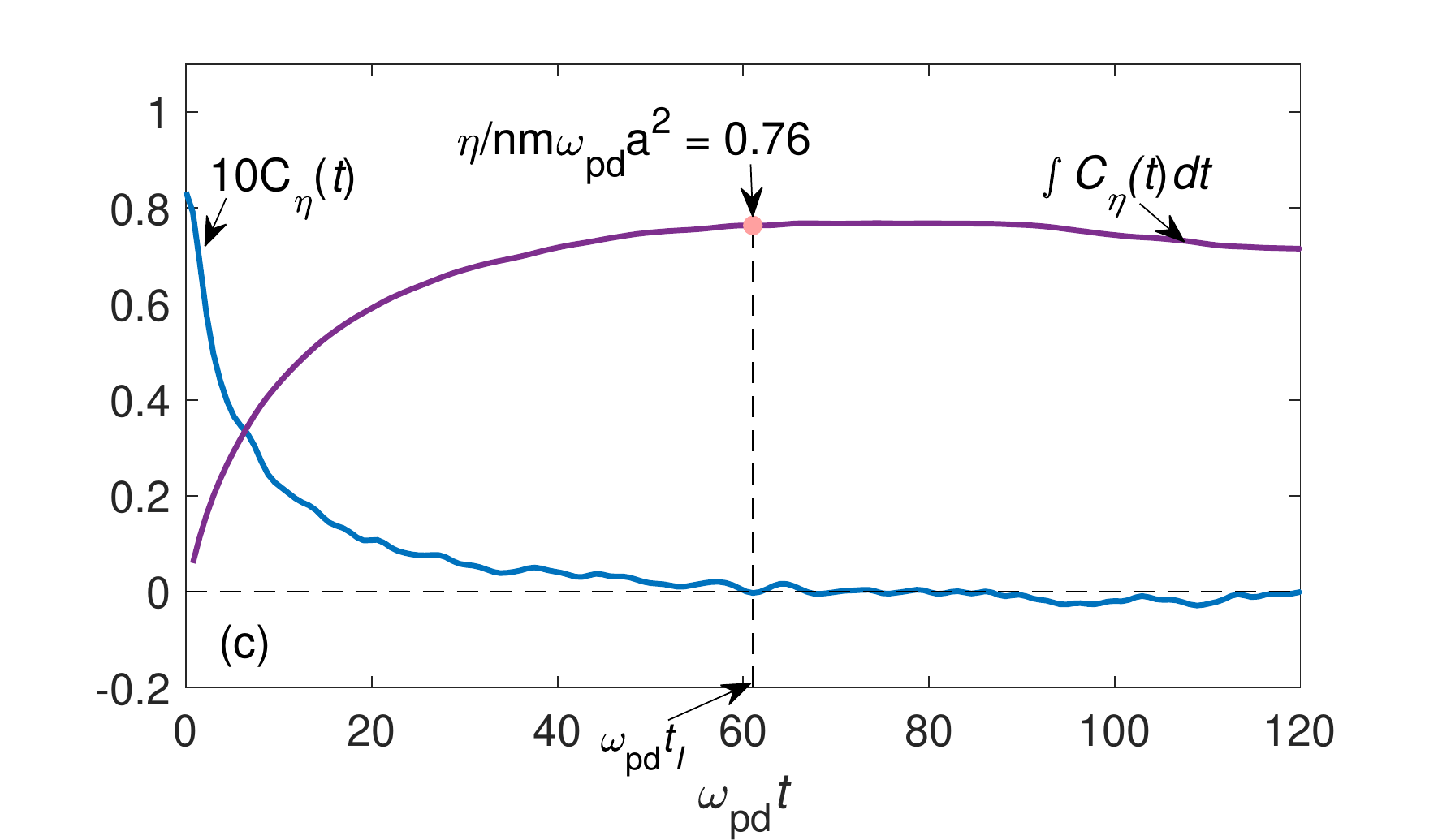}

\caption{\label{figure2}Autocorrelation function of the shear stress $C_{\eta}(t)$ and its time integration for $\Gamma=128.5$. Results are shown three independet simulation runs (a) Run 1, (b) Run2, (c) Run3 with different random number seeds for the initial velocities. The  upper limit of the Green-Kubo integral in Eq.~(\ref{eq4}), $\omega_{pd}t_I$,  is chosen as the time when
this noisy $C_{\eta}(t)$ first crosses zero, indicated by the circle and arrow.$\eta/nm\omega_{pd}a^2$ is the normalized viscosity obtaine from the integral. Note that for better clarity, the autocorrelation function of the shear stress $C_{\eta}(t)$ is multiplied by a factor of 10.}
\end{figure}

\subsection{\label{Viscosity}Shear viscosity}
In dusty plasma studies, the viscosity has been quantified using two methods, with and without macroscopic velocity gradient. In dusty plasma experiments, macroscopic shear stress is applied to generate macroscopic velocity gradient, and viscosity is calculated using the hydrodynamic approach~\cite{Gavrikov2005,Vorona2007,Nosenko2004}. In equilibrium simulations, viscosity is calculated with no macroscopic velocity gradient. Here, we calculate the viscosity using the equilibrium MD simulations and will show that our results for viscosity values are in good agreement with those obtained by non-equilibrium simulations. For equilibrium systems, without gradients, the shear viscosity $\eta$  can be calculated by the Green-Kubo relation~\cite{Feng2011}
\begin{equation}
\eta= \int_0^{\infty}C_{\eta}(t)dt=\frac{1}{Ak_BT} \int_0^{\infty}\langle P_{xy}(t)P_{xy}(0)\rangle dt,
\label{eq4}
\end{equation}
where $C_{\eta}(t)$ is the autocorrelation function of the shear stress, and $A$ is the area of the 2D system.
The off-diagonal element of the stress tensor $P_{xy}(t)$ is given by
\begin{equation}
 P_{xy}(t)=\sum_{i=1}^{N}\left(m_i v_{xi}v_{yi} - \sum_{i}\sum_{j>i}\frac{x_{ij}y_{ij}}{r_{ij}}\frac{\partial\phi_{ij}}{\partial r_{ij}}\right),
\label{eq5}
\end{equation}
where $r_{ij}=\vert \boldsymbol{\mathrm{r}}_i-\boldsymbol{\mathrm{r}}_j \vert$, $m_{i}$ is the mass of the particle $i$, and $v_{xi}$ and $v_{yi}$ are the components $x$ and $y$ of the velocity of the particle $i$, respectively.

The upper limit of the time integral in  Eq.~(\ref{eq4}) must be chosen cautiously. A very long integration time due to the noise in the correlation function $C_{\eta}(t)$ leads to unreliable values for viscosity. Therefore, in practice, the infinite upper limit is replaced by a finite time $t_I$ to avoid the statistical errors in the viscosity estimation. In agreement with Refs.~\cite{Feng2011,Donko2010,Feng2012,Feng2013}, we choose the $t_I$ as the time when this noisy $C_{\eta}(t)$
first crosses zero, as shown in Figs.~2(a), (b), (c). We calculate numerically the integral of Eq.~(\ref{eq4}) by the well-known trapezoidal rule ~\cite{Press1992}, and the results for viscosity are shown in Fig.~3. To improve statistical accuracy, we carry out several independent simulation runs with different random number seeds for the initial velocities, and calculate the standard deviation from the mean (i.e., error bar) of these simulations (Table~\ref{jlab1}).
\begin{table}[!htp]
\caption{\label{jlab1}Results for the mean shear viscosity $\bar{\eta}$ from three independent simulation runs normalized by $nm\omega_{pd}a^2$ along with standard deviation from the mean $\sigma_{M}$.}
\begin{ruledtabular}
\bgroup
\def\arraystretch{1.4}
\begin{tabular}{cccccccc}
$\Gamma$&$\bar{\eta}$&$\sigma_{M}$ \\
\hline
3.62&0.250& 0.009\\
8.77&0.156&0.013\\
18.12&0.152&0.007\\
54.46&0.271&0.010\\
89.39&0.406&0.017\\
106.50&0.52&0.03\\
123.42&0.65&0.04\\
128.50&0.76&0.04\\
135.50&1.18&0.09\\
\end{tabular}
\egroup
\end{ruledtabular}
\end{table}
In Figure.~3, we compare the values of the viscosity coefficient obtained from our simulation with the results obtained from the equilibrium molecular dynamics simulation (EMD) under the same conditions in Ref.~[29] and the results obtained from the non-equilibrium molecular dynamics simulation (NEMD) in Ref.~[28]. In NEMD, an appropriate perturbation $E$ is applied, then, the ensemble average of the resulting flux $\langle J\rangle$ is measured and the ratio of flux and field $E$ gives the viscosity coefficient~\cite{Muller1999}. In the method used in Ref.~[28], cause and effect are reversed in a NEMD simulation, i.e., the flux is imposed and the corresponding field is measured.
Our computed values of $\eta$ are in good agreement with the previously simulated shear viscosity values obtained from EMD and NEMD for 2DYL ~\cite{Liu2005,Donko2006}.

\begin{figure}[!htp]

\includegraphics[width=9cm,height=6.5cm]{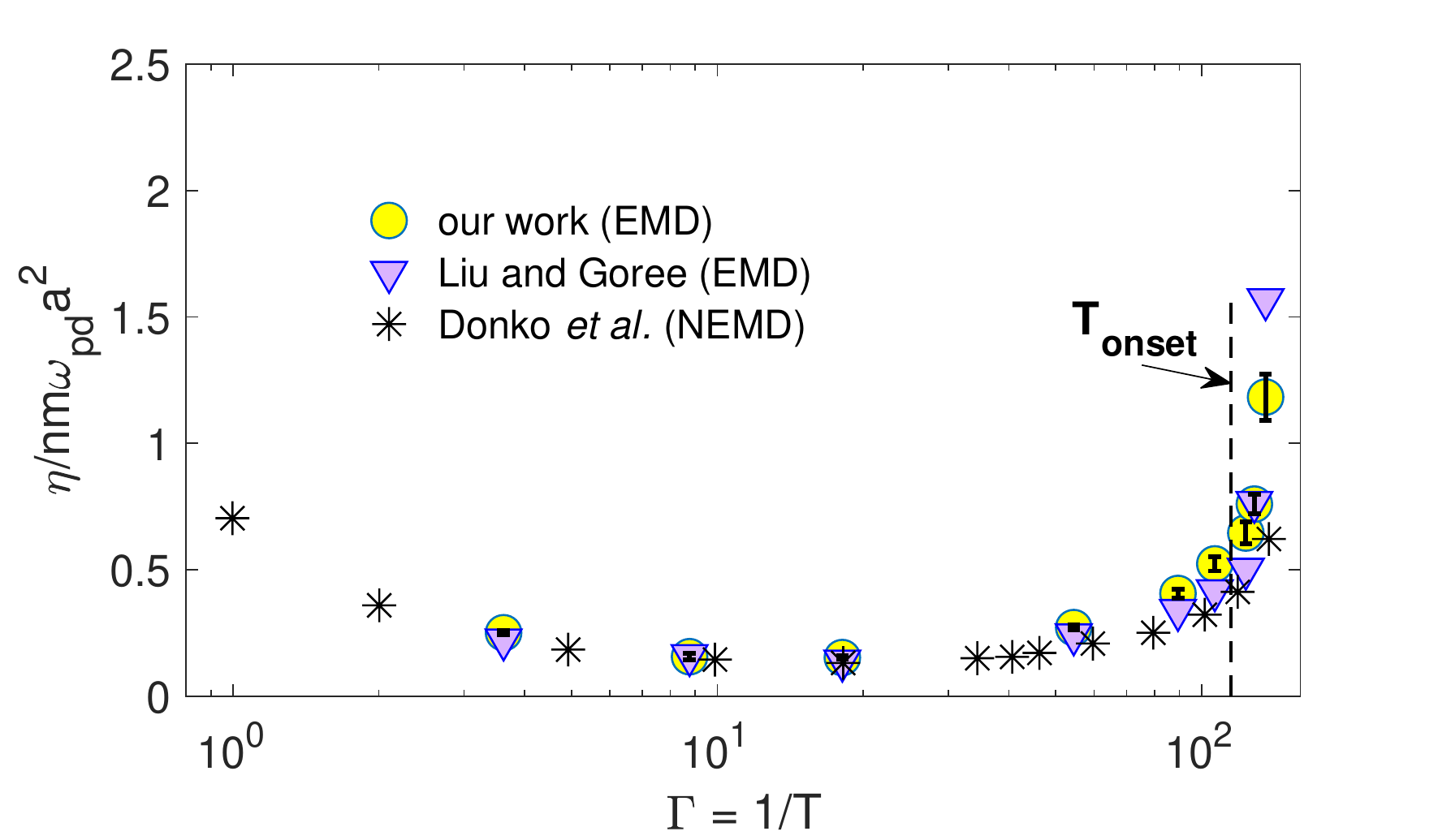}

\caption{\label{figure3} Temperature dependence of the shear viscosity $\eta$. The error bars are calculated from three independent simulation runs. For comparison, data are shown from the equilibrium molecular dynamics simulation (EMD) at $\kappa=0.56$ and $N=1024$~\cite{Liu2005}, and from the non-equilibrium molecular dynamics simulations (NEMD) at $\kappa=0.5$ and $N=1600$~\cite{Donko2006}.}
\end{figure}
Note that in addition to the equilibrium molecular dynamics method, there are other methods for determining the viscosity coefficient, all of which are acceptable within the framework of their assumptions. For example, Haralson and Goree obtained the viscosity by the hydrodynamic method for low temperatures, but as discussed in these papers~\cite{Haralson2016,Haralson2017}, they compared their results with Refs.~[28,29] and found in addition to the similar trend, their viscosity data also
show similar values to simulations. In particular, their experimental viscosity values for $\kappa\approx0.75$ lie between the simulation values for $\kappa=0.5$ and $\kappa=1$ in Ref.~[28].

\begin{figure}[!htp]
\includegraphics[width=9cm,height=6.5cm]{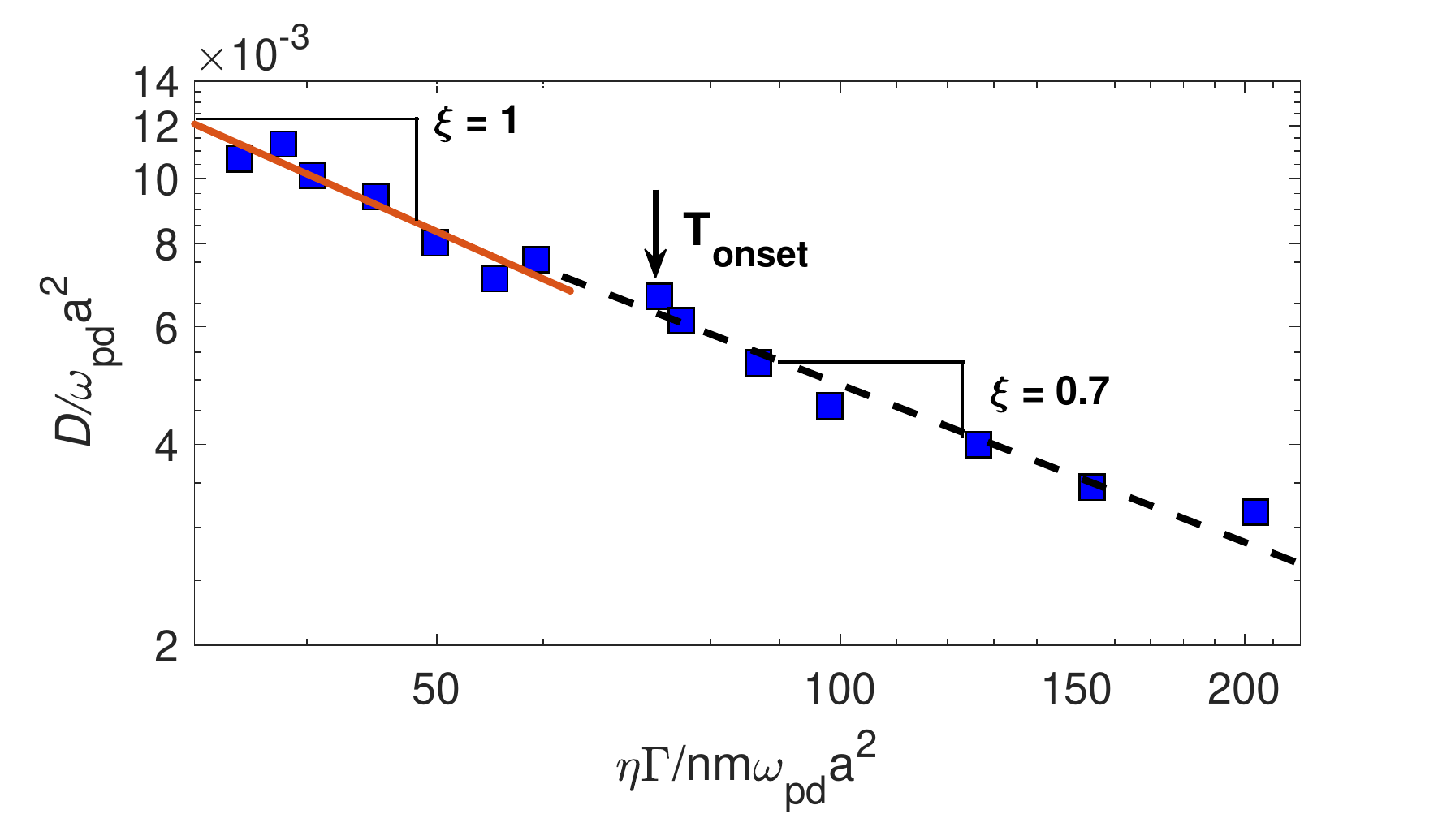}

\caption{\label{figure4}Self-diffusion coefficeint $D$ as the function of shear viscosity scaled by the temperature $\eta\Gamma/nm\omega_{pd}a^2(=\eta/nm\omega_{pd}a^2T)$. The arrow indicates the position of $T_{\rm{onset}}$. The SE relation is obtained as a power-law $D\propto (\eta/T)^{-\xi}$ by the fitting of the data, and $\xi$ is unity for $T>T_{\rm{onset}}$ corresponding to $\Gamma<\Gamma_{\rm{onset}}$ (because $\Gamma=1/T$), satisfying the SE relation. With decreasing temperatre below $T_{\rm{onset}}$ (corresponding to increasing $\Gamma$ above $\Gamma_{\rm{onset}}$ ), the exponent reduces to $\xi=0.7$ indicating the SE violation.}
\end{figure}
The relationship between $D$ and $\eta/T$ is shown in Fig.~4. Data are fitted to a power-law of the form $D\propto (\eta/T)^{-\xi}$. For the temperatures above the onset temperature $T_{\rm{onset}}\approx 0.0087$ (corresponding to $\Gamma_{\rm{onset}} \approx 115$), the exponent $\xi$ is unity indicating the preservation of the SE relation. For temperatures below $T_{\rm{onset}}$ , the exponent is $\xi=0.7$, which is different from unity indicating the violation of the SE relation. The reason for this violation  is explained as follows: With decreasing temperature, the shear viscosity increases but the self-diffusion coefficient decreases (Fig.~1(b)). At temperatures above the $T_{\rm{onset}}$, $\eta$ and $D$ are coupled together, which means that rates for increasing viscosity and decreasing self-diffusion coefficient compensate each other, and the SE relation is preserved. However, with decreasing temperature below the $T_{\rm{onset}}$, the rate of increase in the shear viscosity is greater than the rate of decrease in the self-diffusion coefficient, i.e., the increase in the shear viscosity is faster than the decrease in the self-diffusion coefficient, consequently, this lack of coupling between $\eta$ and $D$ leads to the violation of the SE relation in the 2DYL near the melting point.

\section{\label{Time scales}Time scales}
To answer the question as to what is the origin of the SE violation in 2DYL, we study the time scales that support the violation.
We first focus on the structural relaxation time of the incoherent density-density correlation function or the self-intermediate scattering function $F_s(\boldsymbol{\mathrm{k}},t)$. This function is the Fourier transform of the distribution of the particle displacement,  defined as ~\cite{Hansen}
\begin{equation}
 F_s(\boldsymbol{\mathrm{k}},t)=\frac{1}{N}\Bigg\langle\sum_{\mathit{i}=1}^N \rm{exp}\left(-i\boldsymbol{\mathrm{k}}\cdot[\boldsymbol{\mathrm{r}}_\mathit{i}(t)-\boldsymbol{\mathrm{r}}_\mathit{i}(0)]\right)\Bigg\rangle,
\label{eq6}
\end{equation}
where $\boldsymbol{\mathrm{k}}$ is the wave vector.
For an isotropic system, $F_s(\boldsymbol{\mathrm{k}},t)$ depends only on the magnitude $k= \vert\boldsymbol{\mathrm{k}}\vert$, therefore, averaging over all directions yields~\cite{Ghannad2019}

\begin{equation}
\big\langle\rm{exp}\left(-i\boldsymbol{\mathrm{k}}.\boldsymbol{\mathrm{r}}\right)\big\rangle_{\phi}=\frac{1}{2\pi}\int_0^{2\pi}\rm{exp}\left(-i\mathit{kr} cos\phi\right)\mathit{d}\phi=\mathit{J}_0(\mathit{kr})
\label{eq7}
\end{equation}
where $\phi$ is the angle between the vectors $\boldsymbol{\mathrm{k}}$ and $\boldsymbol{\mathrm{r}}$, and $\mathit{J}_0(\mathit{kr})$ = sin($kr$)/($kr$) is the ordinary Bessel function of order zero. Therefore, for an isotropic system, $F_s(\boldsymbol{\mathrm{k}},t)$ reduces to 
\begin{equation}
F_s(k,t)=\frac{1}{N}\Bigg\langle \sum_{\mathit{i}=1}^N\frac{\mathrm{sin}(k\vert \boldsymbol{\mathrm{r}}_\mathit{i}(t)-\boldsymbol{\mathrm{r}}_\mathit{i}(0)\vert)}{k\vert \boldsymbol{\mathrm{r}}_\mathit{i}(t)-\boldsymbol{\mathrm{r}}_\mathit{i}(0)\vert} \Bigg\rangle.
\label{eq8}
\end{equation} 

We choose the wave number $k= \vert\boldsymbol{\mathrm{k}}\vert=4.23 a^{-1}$, which is the position of the first peak in the static structure factor $S(k)$ of the 2DYL. For an isotropic system in two dimensions, the static structure factor is calculated by~\cite{Pathria}
\begin{equation}
S(k)=1+2\pi n\int_0^\infty r(g(r)-1)J_0(kr)dr\label{eq9}
\end{equation}
where $g(r)$ is the radial distribution function and for a homogeneous uniform system is defined by~\cite{Hail}
\begin{equation}
g(r)=\frac{1}{n}\Big\langle\frac{1}{N}\sum_{i=1}^N\sum_{j\neq i}^N\delta(r-\vert r_i-r_j\vert)\Big\rangle,\label{eq10}
\end{equation}
where $\delta$ is the Dirac delta function. The functions $g(r)$ and $S(k)$ for the 2DYL are shown in Figs.~5(a), (b).

\begin{figure}[!htp]

\includegraphics[width=9cm,height=6.5cm]{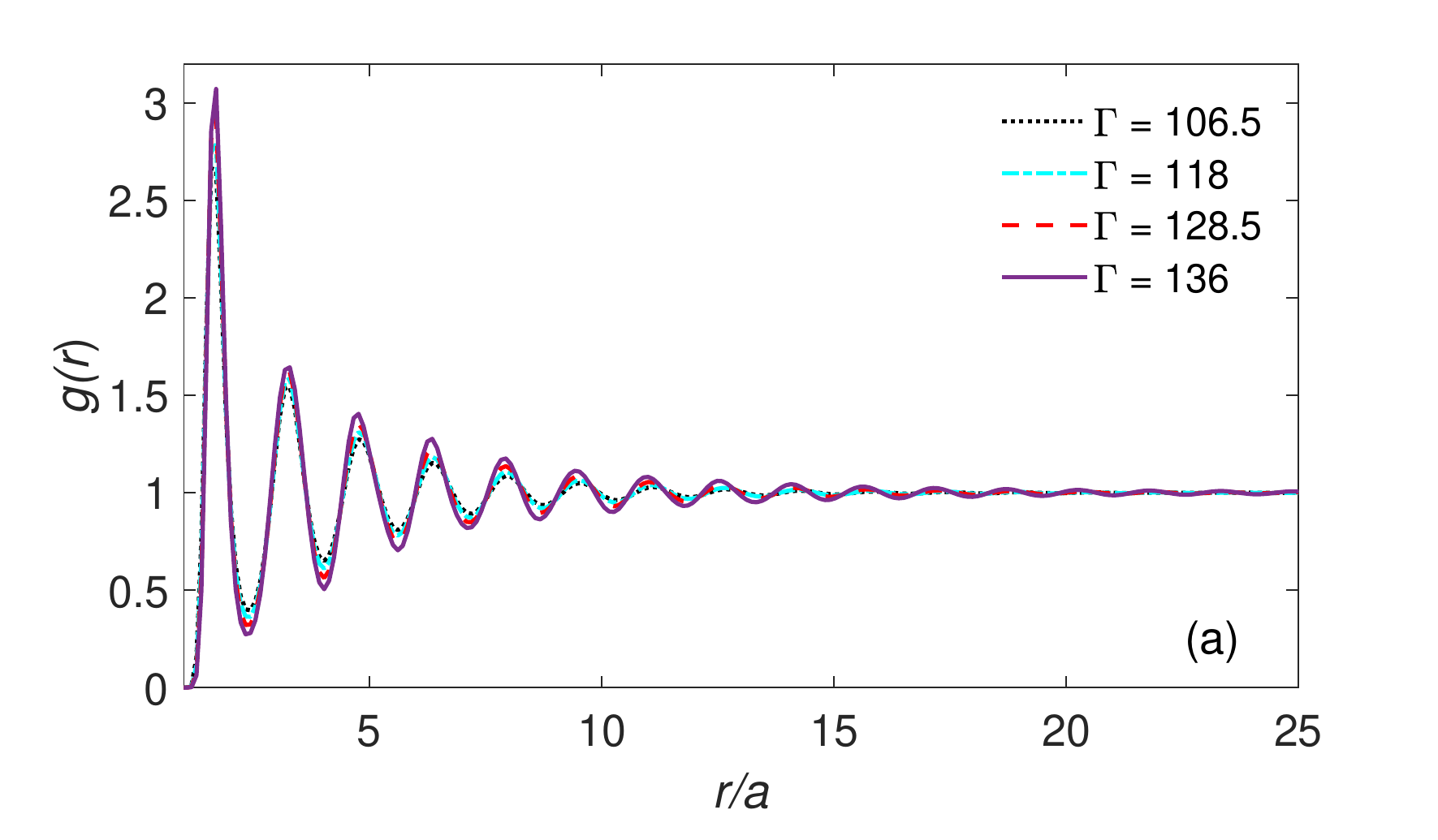}
\includegraphics[width=9cm,height=6.5cm]{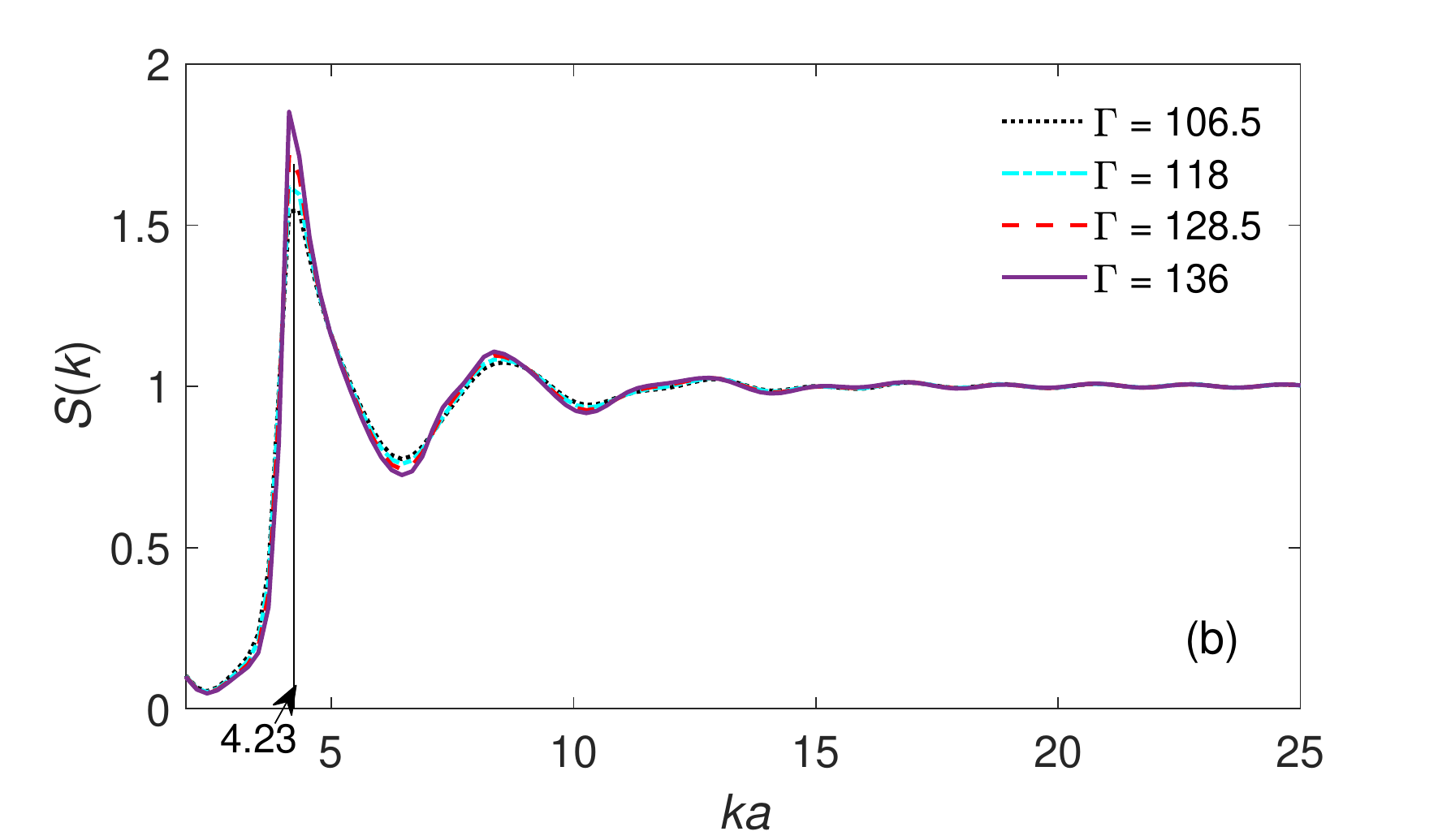}

\caption{\label{figure5} (a) Radial distribution functions $g(r)$ vs distance in units of the
Wigner-Seitz radius $a$ for the 2D Yukawa liquid in various temperatures ($\Gamma$ = 106.5, 118, 128.5, 136). (b) Static structure factor $S(k)$ vs reduced wave number $ka$  in various temperatures. The first peak occurs at $ka$=4.23.}
\end{figure}

In a diffusive regime with linear MSD, $F_s(k,t)$ is obtained theoretically by
\begin{equation}
F^{\mathrm{Gauss}}_s(k,t)=\mathrm{exp}\left(-\frac{k^2\langle(\Delta \mathrm{r}(t))^2\rangle}{4}\right)=
\mathrm{exp}\left(-k^2Dt\right),
\label{eq11}
\end{equation}
which is known as the Gaussian approximation~\cite{Hansen}. Here, MSD = $\langle(\Delta \it{r}(t\rm))^2\rangle$. Deviations  of the particle displacements from a Gaussian distribution are quantified by kurtosis. We used the excess kurtosis or 
non-Gaussian parameter $\alpha_2(t)$, as~\cite{Rahman1964,Charbonneau2012}
\begin{equation}
\alpha_2(t)=\frac{d}{d+2}\frac{\langle(\Delta \mathit{r}(t))^4\rangle}{\langle(\Delta \mathit{r}(t))^2\rangle^2}-1,
\label{eq12}
\end{equation}
where $d$ denotes spatial dimension, and For two dimensions, it reduces to 
\begin{equation}
\alpha_2(t)=\frac{1}{2}\frac{\langle(\Delta \mathit{r}(t))^4\rangle}{\langle(\Delta \mathit{r}(t))^2\rangle^2}-1.
\label{eq13}
\end{equation} 
For a Gaussian distribution, $\alpha_2(t)$ approaches zero.

\begin{figure}[!htp]

\includegraphics[width=9cm,height=6.5cm]{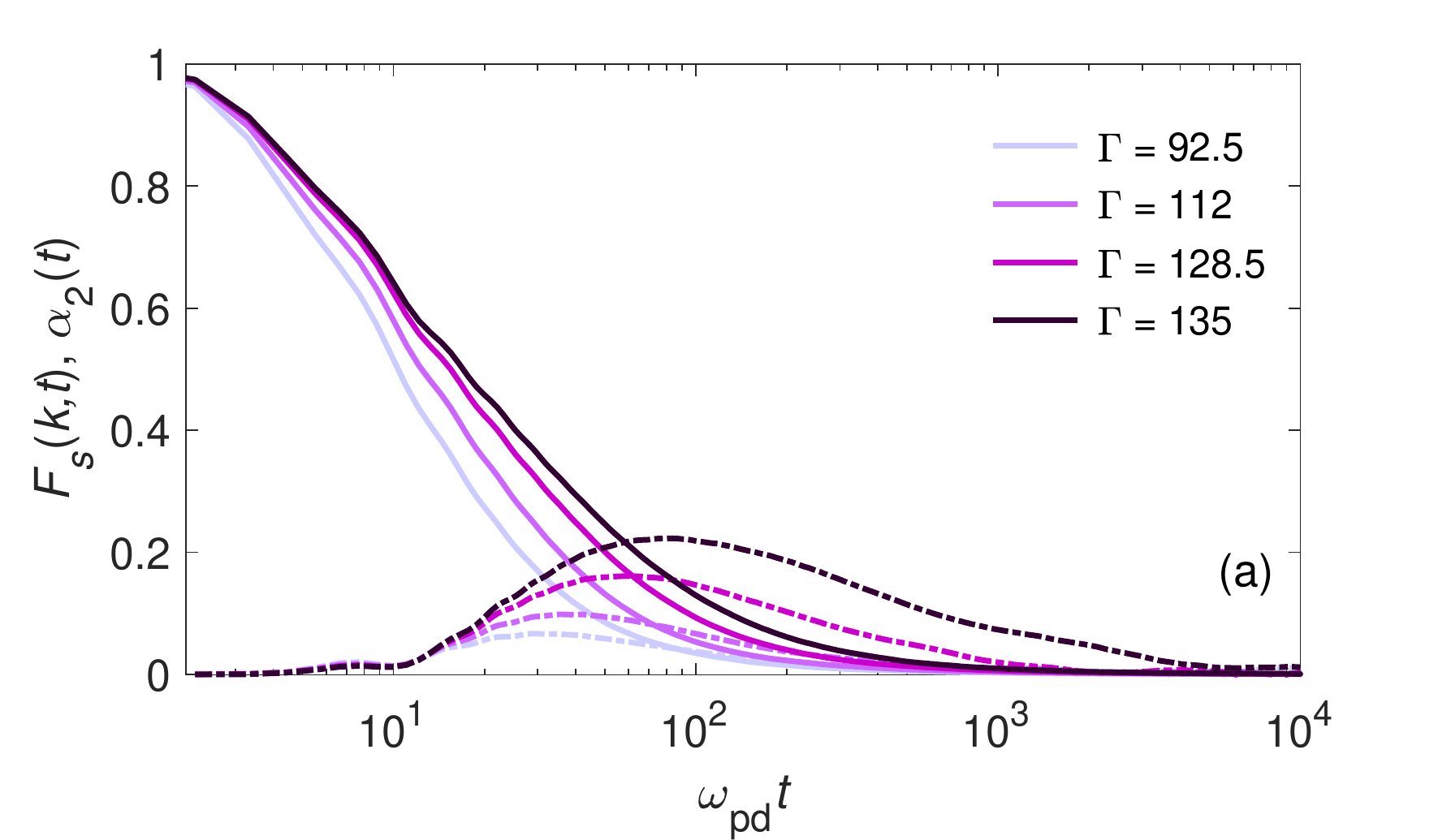}

\includegraphics[width=9cm,height=6.5cm]{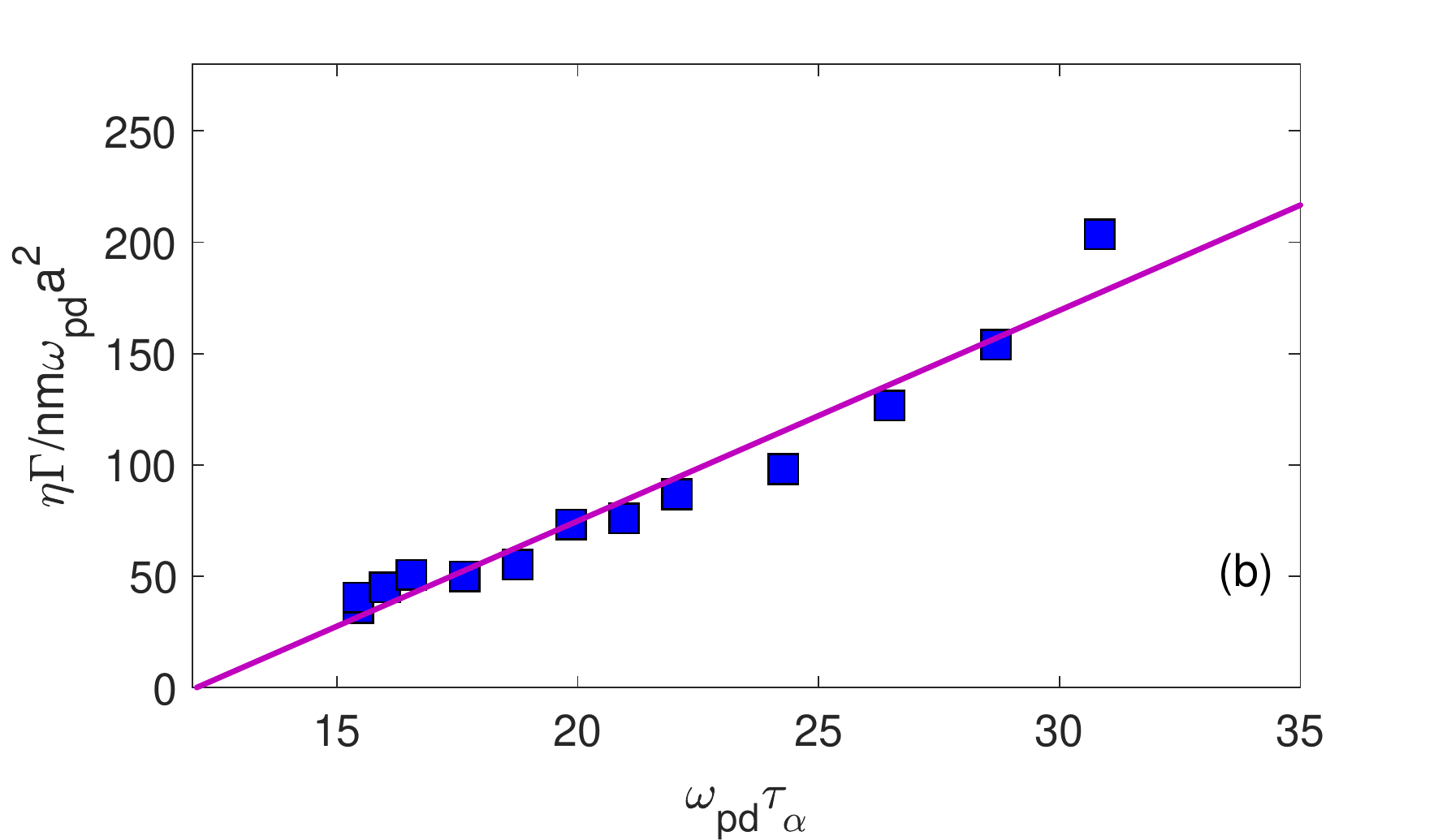}

\includegraphics[width=9cm,height=6.5cm]{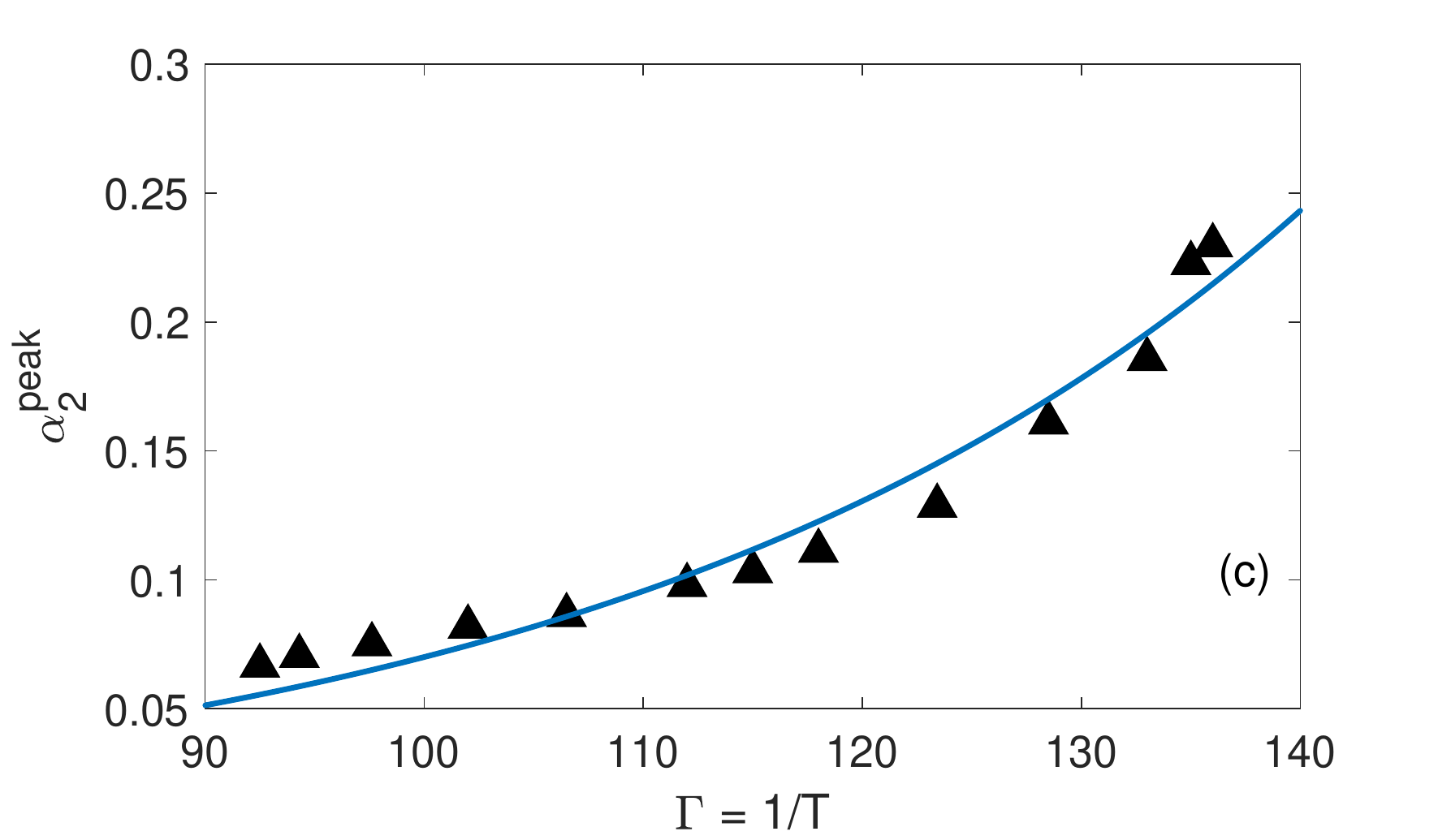}

\caption{\label{figure6} (a) The self-intermediate scattering function $F_s(k,t)$ (solid lines) as well as the non-Gaussian parameter $\alpha_2(t)$ (dash-dotted lines). (b) Relationship between $\eta/T(=\eta\Gamma)$ and $\tau_{\alpha}$. Solid line is a linear fit
to the simulation results. (c) Temperature dependence of the peak height $\alpha^{\rm{peak}}_2$ of the non-Gaussian parameter $\alpha_2(t)$. The peak hights of $\alpha_2(t)$ increase with decreasing temperature.The solid line a fit to guide the eye.}
\end{figure}

As shown in Fig.~6(a), for $\Gamma=92.5$, $\alpha_2(t)$ is nearly zero and it increases with decreasing temperature ($=$ increasing $\Gamma$) indicating  the non-Gaussian behavior of the function $F_s(k,t)$ in the 2DYL when cools to near the melting point.

The structure relaxation time $\tau_{\alpha}$ is defined as ~\cite{Becker2006,Jeong2010} the time where the function $F_s(k,t)$ decays to a value of 1/e. If $\tau_{\alpha}$ is proportional to $\eta/T$, we have another form for SE relation as $D\propto \tau_{\alpha}^{-1}$. The justification for replacing $\tau_{\alpha}$ with $\eta/T$ results from Eq.~(\ref{eq11}), i.e, $F^{\rm{Gauss}}_s(k,t)=\mathrm{exp}(-k^2Dt)\equiv \mathrm{exp}(-t/\tau)$, consequently, $\tau^{-1}$=$k^2D$=$\frac{k_Bk^2}{c'\pi}\frac{T}{\eta}$. As shown in Fig.~6(b) , the proportional relationship $\eta/T\propto\tau_{\alpha}$ holds in 2DYL. Therefore, the relation $D\tau_{\alpha} = constant$, can be applied as an alternative for the SE relation. With decreasing temperature, the SE relation is violated and the peak hights of $\alpha_2(t)$ increase (Fig.~6(c)). We denote the peak time of $\alpha_2(t)$ with $\tau_{\alpha_2}$, i.e., when the non-Gaussian parameter reaches its maximum value. The results obtained from our simulation data show a linear relationship between the $\tau_{\alpha_2}$ and $\tau_{\alpha}$ (Fig.~7) indicating that the relation $D\tau_{\alpha_2} = costant$ is another acceptable form of the SE relation.

The temperature dependence of $D\tau_{\alpha}$ and $D\tau_{\alpha_2}$ along with that of $D\eta/T$ is shown in Fig.~8. At temperatures much higher than the phase transition temperature, the SE relations, $D\eta/T, D\tau_{\alpha}$, and $D\tau_{\alpha_2}$, are preserved. However, with decreasing temperature to near
the phase transition temperature, the SE relations are violated. As a result, the SE violation is linked with deviations of the particle displacements from Gaussian.

\begin{figure}[!htp]

\includegraphics[width=9cm,height=6.5cm]{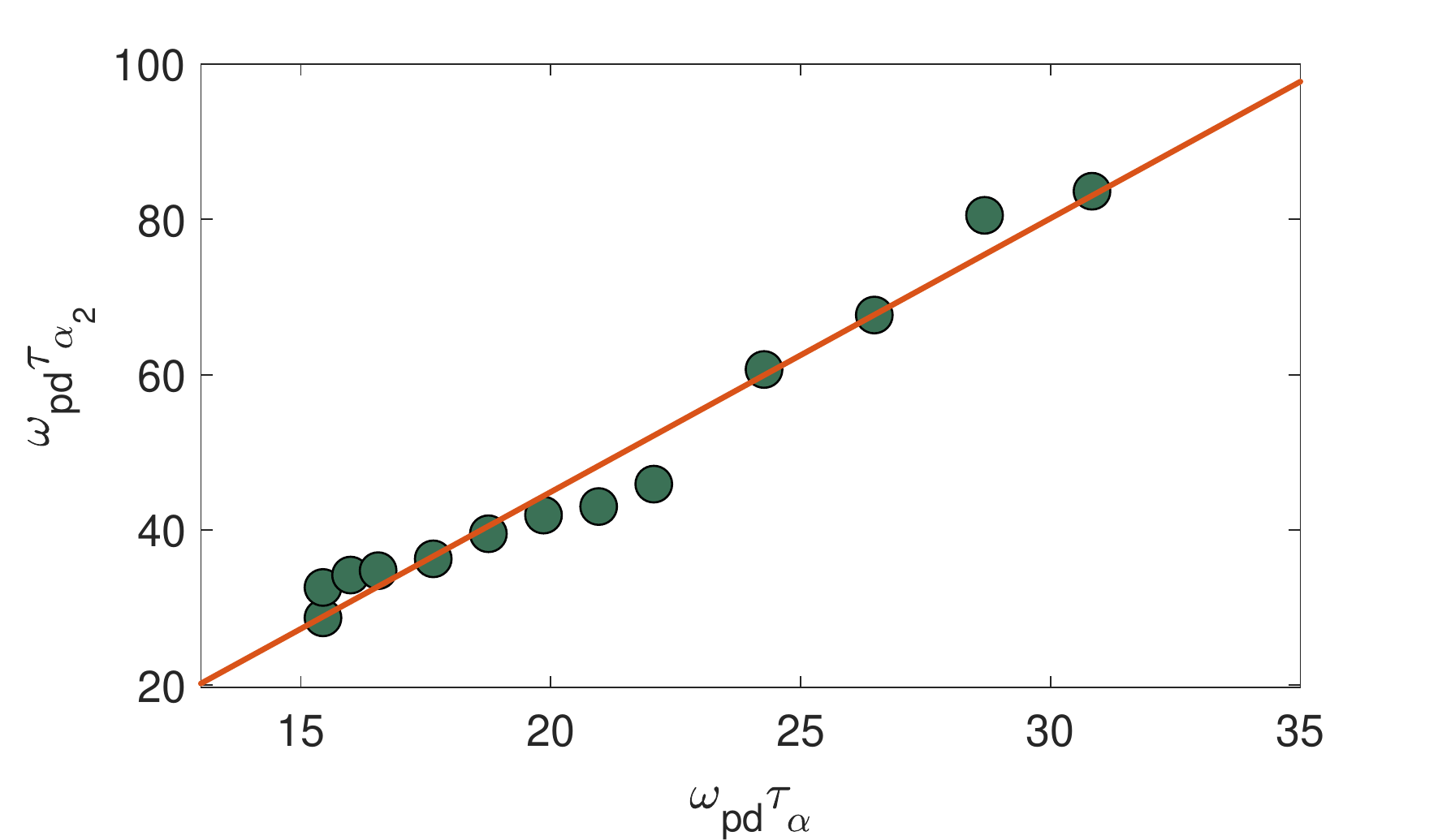}

\caption{\label{figure7}Relationship between $\tau_{\alpha}$ and $\tau_{\alpha_2}$.  Solid line is a linear fit
to the simulation results.}
\end{figure}

\begin{figure}[!htp]

\includegraphics[width=9cm,height=6.5cm]{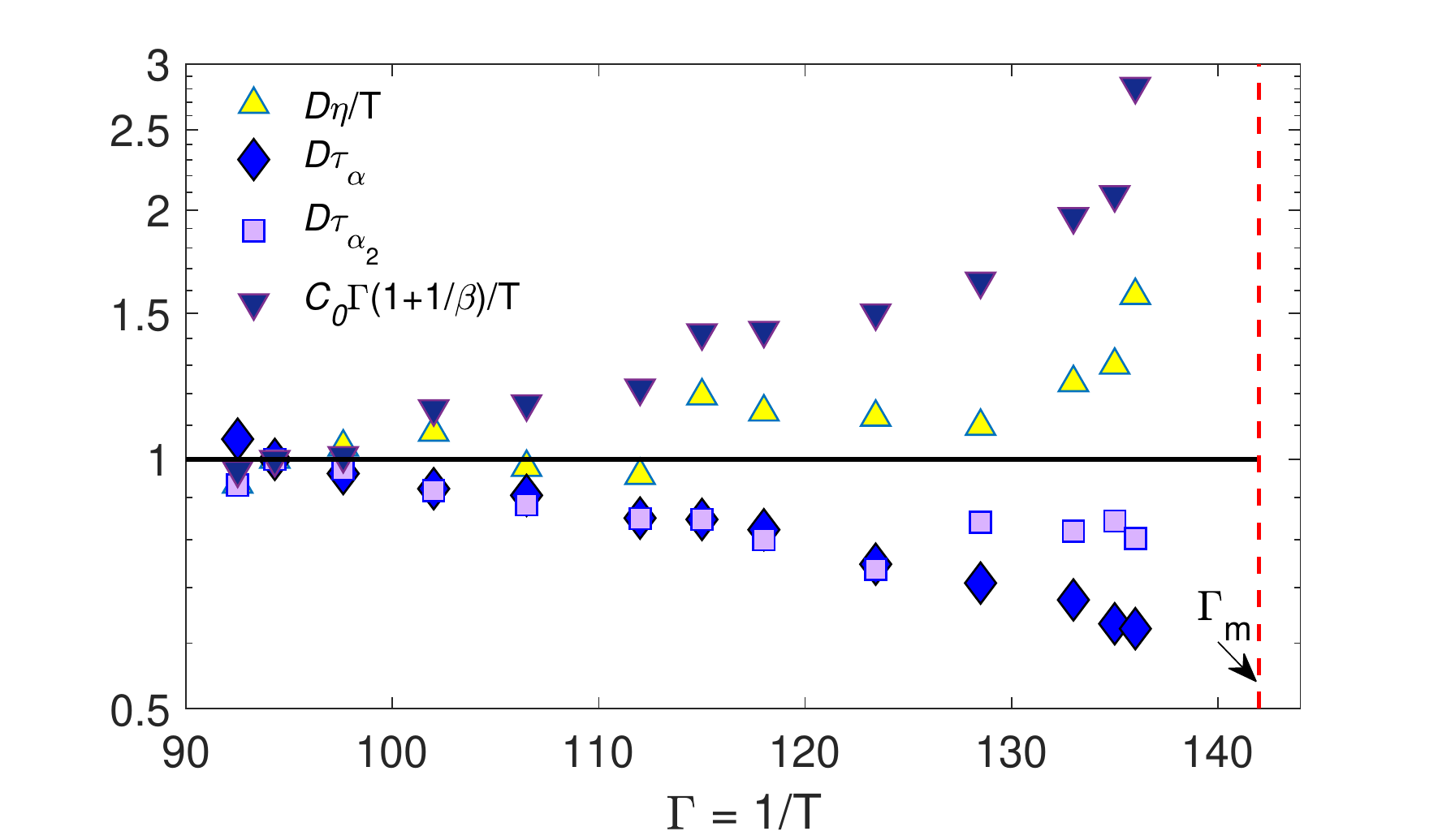}

\caption{\label{figure8}Temperature dependence of the SE relations $D\eta/T, D\tau_{\alpha}, D\tau_{\alpha_2},$ and $C_0\Gamma(1+1/\beta)/T$ normalized to their values at $T = 0.0106$ $ (\Gamma = 94.3)$. All the SE relations indicate the SE violation with desreasing temperature to near the phase transition temperature $T_m$ corresponding to increasing $\Gamma$ to near $\Gamma_m$.}
\end{figure}

\begin{figure}[!htp]

\includegraphics[width=9cm,height=6.5cm]{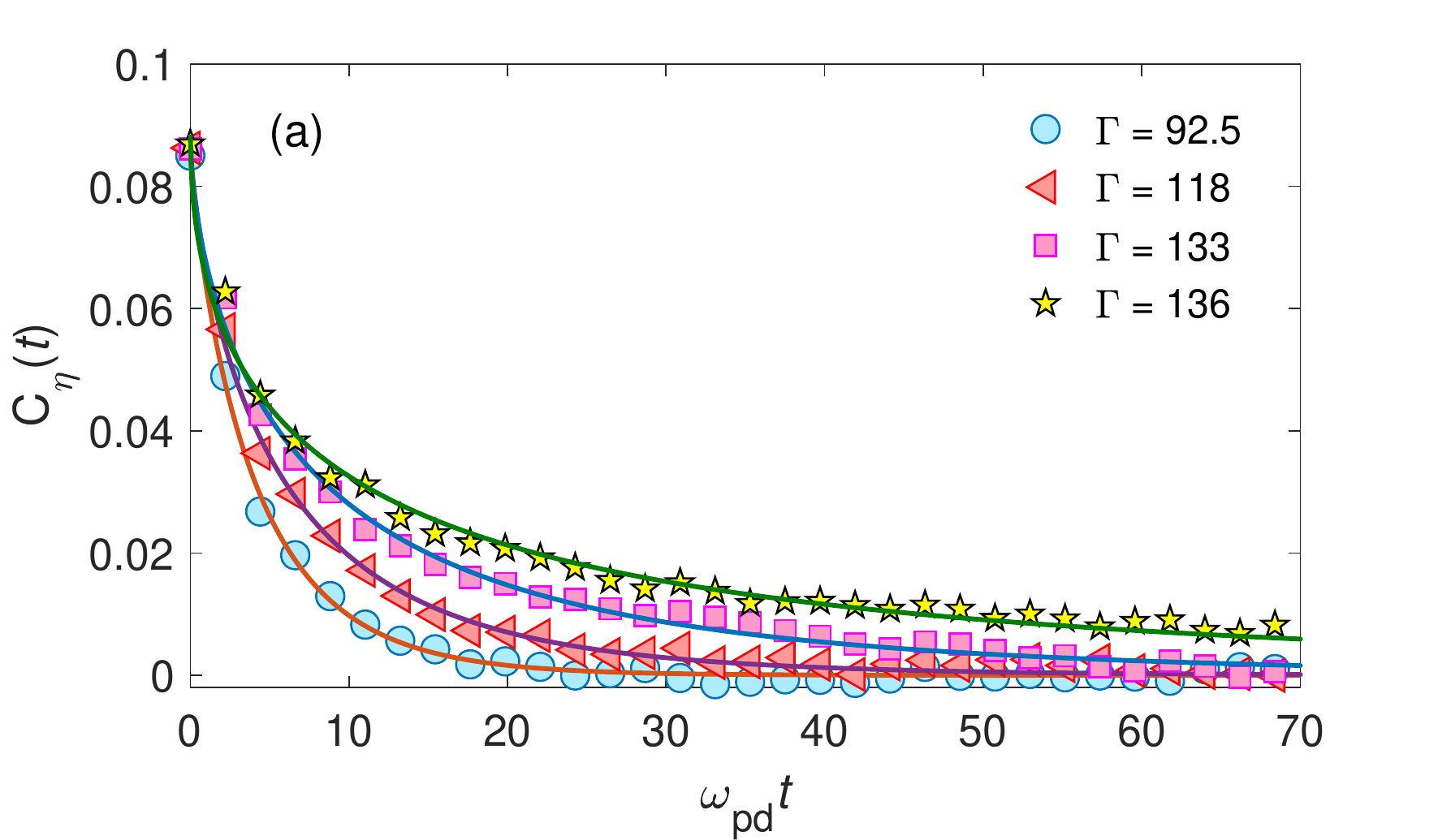}

\includegraphics[width=9cm,height=6.5cm]{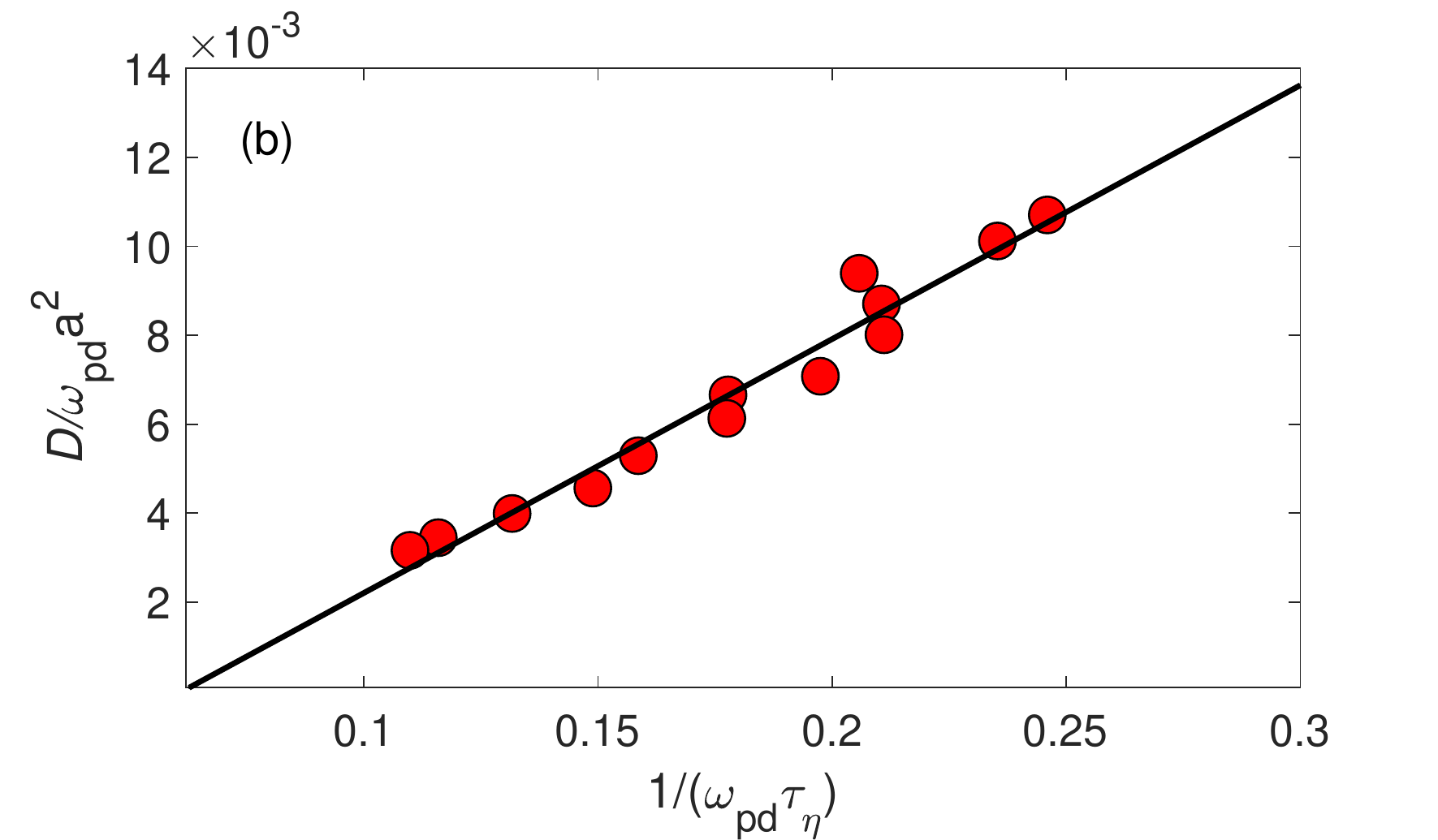}

\includegraphics[width=9cm,height=6.5cm]{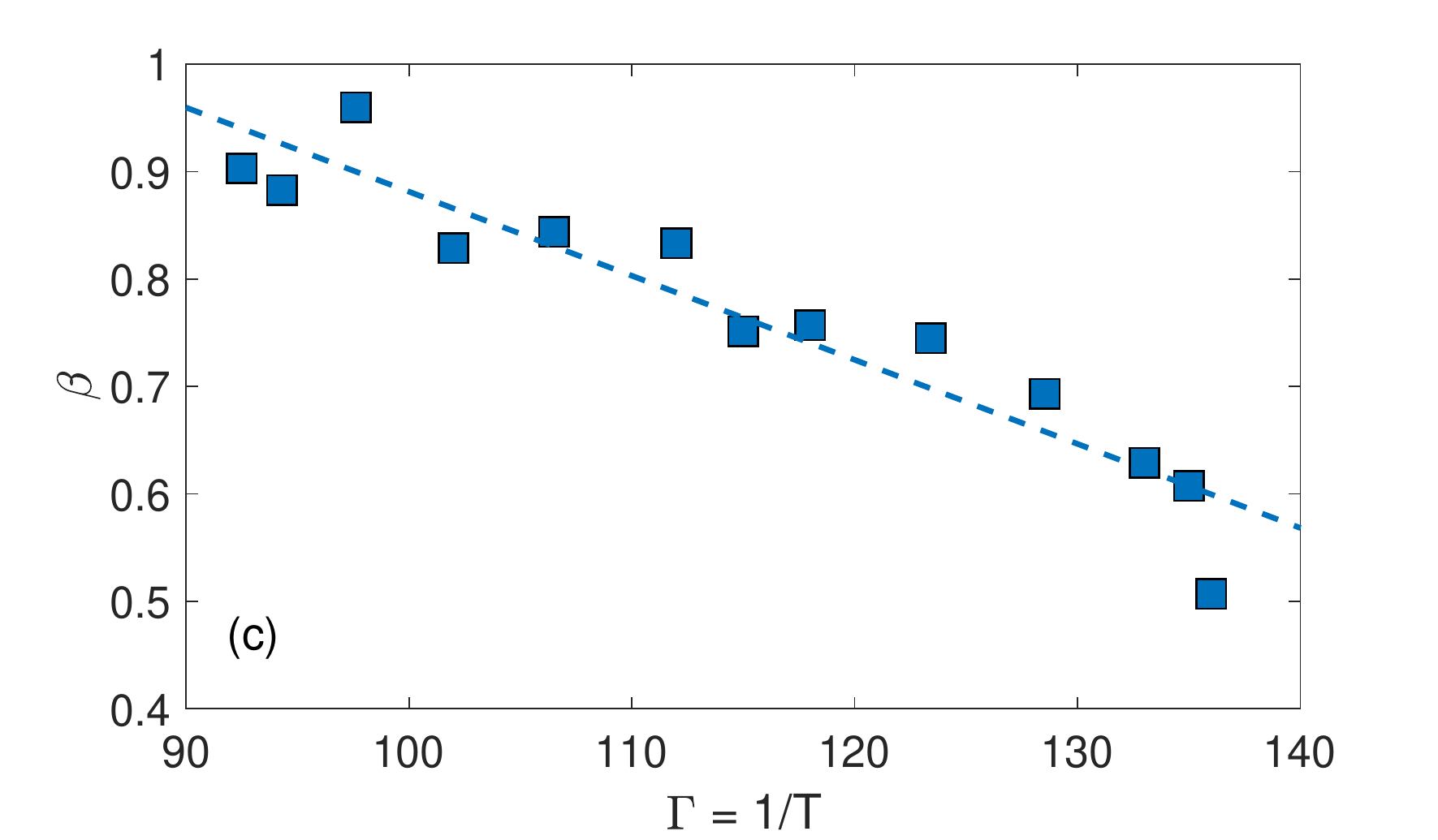}

\caption{\label{figure9}
 (a) Time dependence of the autocorrelation function of the shear stress $C_{\eta}(t)$ fitted by  a stretched exponential function (solid line). (b) Relationship between $D$ and $\tau_{\eta}^{-1}$.  Solid line is a linear fit to the simulation results. (c) Temperature dependence of the stretched exponent $\beta$. Decreasing $\beta$ with decreasing temperature implies increasing the non-exponentiality of the shear stress autocorrelation function. The dashed line is a guide to the eye.}
\end{figure}

Next, we look at the time dependence of the autocorrelation function of the shear stress. It is well fitted by a stretched exponential function $C_{\eta}(t)\simeq C_0 \mathrm{exp}(-(t/\tau_{\eta})^{\beta})$ where $C_0$ is the amplitude of the stretched exponential, $\tau_{\eta}$ is the stress decay time, and the exponent $\beta$, ranging from 0 to 1, is the degree of non-exponentiality, which determines the degree of deviations from an exponential (Fig.~9(a)). All three parameters are given by fitting. Then, the shear viscosity can be approximated by substituting the stretched exponential into Eq.~(\ref{eq4}) as
\begin{equation}
\eta\approx \int_0^{\infty}C_0 e^{-(t/\tau_{\eta})^{\beta}} dt =\frac{C_0\tau_{\eta}}{\beta}\Gamma(1/\beta)= C_0\tau_{\eta}\Gamma(1+1/\beta),
\label{eq14}
\end{equation}
where $\Gamma$ is the Gamma function.
\begin{figure}

\includegraphics[width=9cm,height=6.5cm]{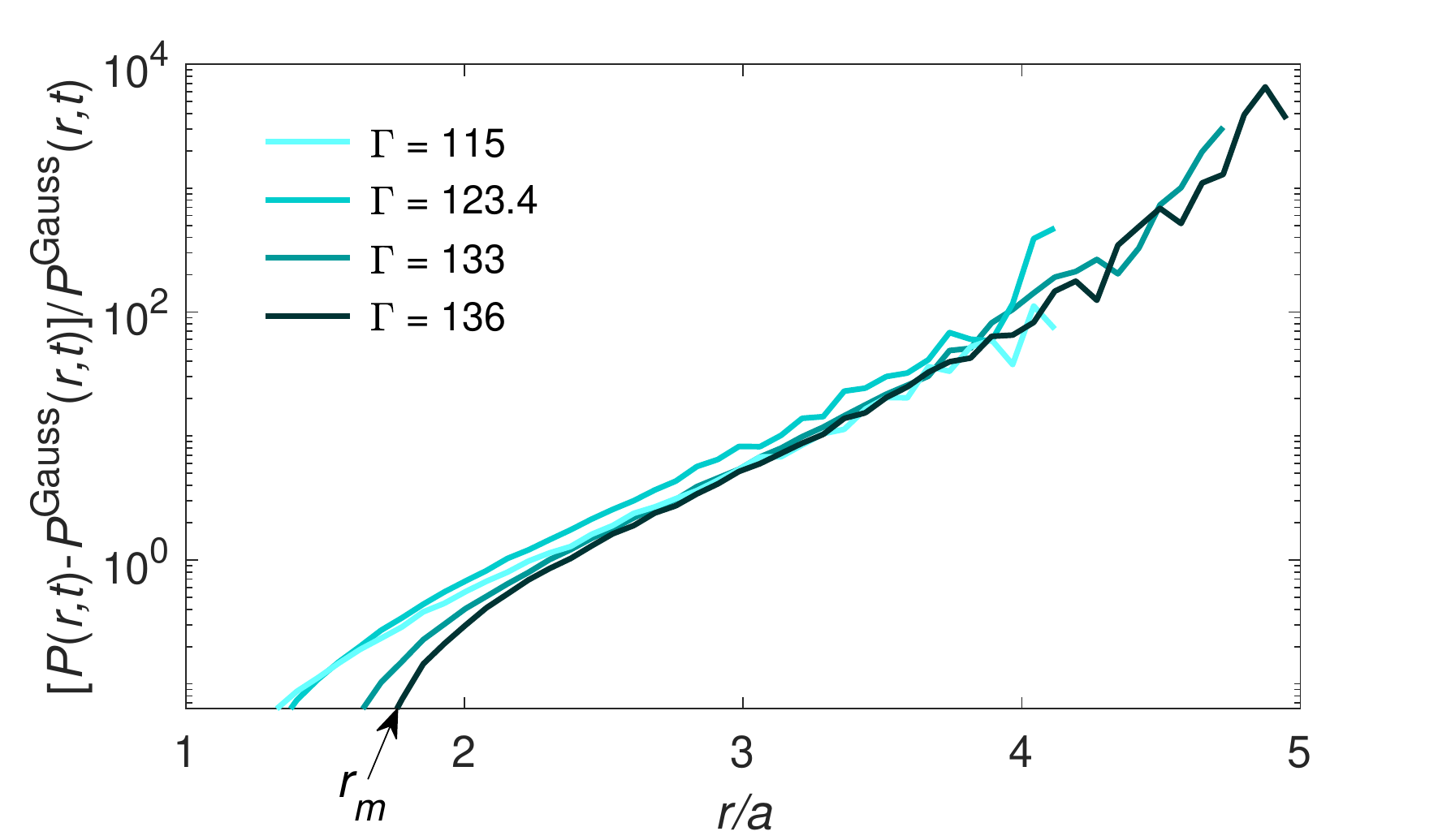}

\caption{\label{figure10}
Relative differences between $P(r,t)$ and Gaussian distribution $P^{\rm{Gauss}}(r,t)$ vs normalized displacement $r/a$ at time $\omega_{pd}\tau_{\alpha_2}$. For $r<r_m$, the relative differences are nearly zero.}
\end{figure}

The relationship between $D$ and inverse $\tau_{\eta}$ is shown in Fig.~9(b).
Our simulation data show a proportional relationship between $D$ and $\tau_{\eta}^{-1}$ for 2DYL. As a result, by using $D\propto\tau_{\eta}^{-1}$ and Eq.~(\ref{eq4}) for $\eta$, we obtain another acceptable form for the SE relation as $D\eta/T\propto C_0\Gamma(1+1/\beta)/T$ shown in Fig.~8, that confirms the SE violation with decreasing temperature. The temperature dependence of $\beta$ is shown in Fig.~9(c). With decreasing temperature, $\beta$ decreases, i.e., deviations from exponentiality of the autocorrelation function of the shear stress increases. Therefore, the SE violation is linked with deviations of the shear stress autocorrelation function $C_{\eta}(t)$ from exponential shape.

Overall, we find that the physical origins of the SE violation in 2DYL are attributed to the non-exponentiality of the stress correlation function and the non-Gaussianity of the distribution of particle displacements. The smaller exponent $\beta$, the greater the deviation from the exponential, which results in stronger heterogeneous dynamics.
The relationship between the non-Gaussian behavior of the distribution of particle displacements and dynamic heterogeneity can also be explained by calculating the relative difference between the distribution of particle displacements $P(r,t)$ and that obtained from the Gaussian approximation $P^{\rm{Gauss}}(r,t)$ given by $P^{\rm{Gauss}}\rm{(}\mathit{r},\it{t}\rm{)}=(1/\pi\langle(\Delta \mathit{r}(\mathit{t}))^2\rangle)\rm{exp}(-\it{r}\rm^2/\langle(\Delta \mathit{r}(\it{t}\rm))^2\rangle)$, where $\langle(\Delta \mathit{r}(t))^2\rangle$ is obtained from the simulation. 
The distribution of particle displacements $P(\boldsymbol{\mathrm{r}},t)$ is given by~\cite{Hail}
\begin{equation}
P(\boldsymbol{\mathrm{r}},t)=\frac{1}{N}\Bigg\langle\sum_{\mathit{i}=1}^N\delta(\boldsymbol{\mathrm{r}}-\boldsymbol{\mathrm{r}}_i(t)+\boldsymbol{\mathrm{r}}_i(0))\Bigg\rangle,
\label{eq15}
\end{equation}
For isotropic liquids, $P(\boldsymbol{\mathrm{r}},t)$ depends only on the scalar  
distance, $r=\vert\boldsymbol{\mathrm{r}}\vert$. Thus, Eq.~(\ref{eq15}) reduces to
\begin{equation}
P(r,t)=\frac{1}{N}\Bigg\langle\sum_{\mathit{i}=1}^N\delta(r-\vert r_i(t)-r_i(0)\vert)\Bigg\rangle
\label{eq16}.
\end{equation}
Physically, 2$\pi rP(r,t)dr$ measures the probability of finding a dust particle at distance $r$ 
from an origin at time $t$ given that the same dust particle was at the origin at the initial time $t=0$~\cite{Hail}. Thus, $P(r,t)$ is normalized by
\begin{equation}
\int P(r,t)d\boldsymbol{\mathrm{r}}=1
\label{eq17}.
\end{equation}

The relative differencs between $P\rm{(}\it{r},t\rm{)}$ and $P^{\rm{Gauss}}\rm{(}\it{r},t\rm{)}$ are shown in Fig.~10, for $t=\tau_{\alpha_2}$. We first note that with decreasing temperature, the relative difference increases, which supports the violation of the SE with decreasing temperature corresponding to increasing $\Gamma$.
For displacements smaller than $r_m\simeq 1.7a$, the relative difference between $P(r,t)$ and $P^{\rm{Gauss}}(r,t)$ is nearly zero, i.e., $P(r_m,\tau_{\alpha_2})=P^{\rm{Gauss}}(r_m,\tau_{\alpha_2})$. However, for $r>r_m$, the relative difference increases and becomes as large as $10^3$, which belongs to the lowest temperature, i.e., $T=0.00735$ or $\Gamma=136$. It means that a significant number of particles travel farther than $r_m$ at time $\omega_{pd}\tau_{\alpha_2}$. It reflects the existence of dynamic heterogeneity with decreasing temperature, i.e., the regions are formed in 2DYL that dust particles are more mobile than expected from a Gaussian approximation.
\section{\label{Conclusions}CONCLUSIONS}
We have reported extensive molecular dynamics results to recognize the origin of the Stokes-Einstein violation in 2D Yukawa liquids.
First, we have computed the mean-squared displacement MSD($t$) at different temperature ranges in the Yukawa liquid. Then, for the temperature ranges that  MSD($t$) is linear and therefore the self-diffusion coefficient $D$ is meaningful, we have computed $D$ from the linear fit of the MSD($t$) versus time. Then, we have calculated the shear viscosity using the Green-Kubo integral  with the assumption that there is no macroscopic velocity gradient in the equilibrium. We have found that there is a temperature threshold $T_{\rm{onset}}$ in which the Stokes-Einstein relation, $D\propto (\eta/T)^{-1}$, is preserved for $T>T_{\rm{onset}}$ but is violated for $T<T_{\rm{onset}}$.

To identify the time scales supporting this violation, we have calculated the structural relaxation time $\tau_{\alpha}$ from the incoherent density-density correlation function or the self-intermediate scattering function
$F_s(k,t)$ when decays to a value of $\rm{e^{-1}}$. We have shown that $\tau_{\alpha}$ is proportional to $\eta/T$. As a result,  
We have defined that an alternative relationship to the Stokes-Einstein relation as $D\propto \tau_{\alpha}^{-1}$. Then, by calculating the peak time of deviations of the particle displacements from a Gaussian distribution, $\tau_{\alpha_2}$, we have shown that there is a linear relation between $\tau_{\alpha_2}$ and $\tau_{\alpha}$ indicating that the relation $D\propto \tau_{\alpha_2}^{-1}$
is another acceptable form of the SE relation. Then, we have shown that with decreasing temperature to below
the phase transition temperature $\Gamma_m$, $D\tau_{\alpha}$ and $D\tau_{\alpha_2}$ are violated which implies that the SE violation is linked with the non-Gaussian behavior of the particle displacements distribution. With calculation $\eta$ from the autocorrelation function of the shear stress approximated by the stretched exponential, $C_{\eta}\propto \mathrm{exp}(-(t/\tau_{\eta})^{\beta})$, we have found another SE relation, and have shown that with decreasing temperature to below $\Gamma_m$, $\beta$ decreases, i.e., deviation from exponentiality of $C_{\eta}(t)$ increases, indicating a link between deviations of the autocorrelation function of the shear stress from the exponential shape and the SE relation.

Generally, our results provide a deep insight into understanding the breakdown of the SE relation, not only in 2D Yukawa liquids but also in other strongly coupled systems such as one-component Coulomb liquids and 3D Yukawa liquids. For these 3D systems, failure of the SE relation has been reported at high temperatures, in contrast to 2D Yukawa liquids~\cite{Daligault2006,Donko2008}. Therefore, it will be interesting to examine the effect of the spatial dimension on the origin of the SE violation.

\section*{\label{DATA AVAILABILITY}DATA AVAILABILITY}
The data that support the findings of this study are available from the author upon reasonable request.
\nocite{*}
\section*{\label{REFERENCES}REFERENCES}
\bibliography{Ghannad}
\end{document}